\documentclass[lettersize,journal]{IEEEtran}
\usepackage{amsmath,amsfonts}
\usepackage{algorithmic}
\usepackage{array}
\usepackage[caption=false,font=normalsize,labelfont=sf,textfont=sf]{subfig}
\usepackage{textcomp}
\usepackage{stfloats}
\usepackage{url}
\usepackage{verbatim}
\usepackage{graphicx}
\usepackage{cite}
\usepackage[ruled,linesnumbered]{algorithm2e}
\usepackage{cleveref}
\usepackage{array}
\usepackage{xcolor}
\usepackage{longtable}
\usepackage{graphicx}
\usepackage{wrapfig}
\usepackage{units}

\begin{document}

\title{Test Case Generation and Test Oracle Support \\ for Testing CPSs using Hybrid Models}

\author{Zahra Sadri-Moshkenani, Justin Bradley, Gregg Rothermel
\thanks{Zahra Sadri-Moshkenani and Gregg Rothermel are with the Department of Computer Science,
North Carolina State University. E-mail:\{zsadrim,gerother\}@ncsu.edu. Justin Bradley is with
the School of Computer Science and Engineering, University of Nebraska-Lincoln, justin.bradley@unl.edu.

This work has been submitted to the IEEE for possible publication. Copyright may be transferred without notice, after which this version may no longer be accessible.}}

\markboth{IEEE TRANSACTIONS ON SOFTWARE ENGINEERING, September 2023} {Sadri-Moshkenani,\MakeLowercase{\textit{(et al.)}:Test Case Generation and Test Oracle Support for Testing CPSs using Hybrid Models}}

\maketitle

\begin{abstract}

Cyber-Physical Systems (CPSs) play a central role in the
behavior of a wide range of autonomous physical systems such
as medical devices, autonomous vehicles, and smart homes
-- many of which are safety-critical.
CPSs are often specified iteratively as a sequence
of models at different levels that can be tested
via simulation systems at early stages of their development cycle.   
One such model is a hybrid automaton; these are used frequently
for CPS applications and have the advantage of
encapsulating both continuous and discrete CPS behaviors.
When testing CPSs, engineers can take advantage
of these models to generate test cases that
target both types of these behaviors.
Moreover, since these models are constructed
early in the development process for CPSs, they
allow test cases to be generated early in that
process for those CPSs -- even before simulation
models of the CPSs have been designed. 
One challenge when testing CPSs is that these 
systems may operate differently even under 
an identically applied test scenario. 
In such cases, we cannot employ test oracles
that use predetermined deterministic behaviors;
instead, test oracles should consider sets of 
desired behaviors in order to determine whether
the CPS has behaved appropriately.
In this paper we present a test case generation
technique, {\sc HyTest}, that generates test
cases based on hybrid models, accompanied by
appropriate test oracles, for use in testing 
CPSs early in their development cycle.
To evaluate the effectiveness and efficiency of 
{\sc HyTest}, we conducted an empirical study in which 
we applied the technique to several CPSs and measured 
its ability to detect faults in those CPSs and the 
amount of time required to perform the testing process. 
The results of the study show that {\sc HyTest} was 
able to detect faults more effectively and efficiently
than the baseline techniques we compare it to.

\end{abstract}


\section{Introduction}
\label{sec:intro}

\noindent \IEEEPARstart{C}{YBER-PHYSICAL} Systems (CPSs) involve a set of integrated 
software components and hardware devices that communicate with one another and 
interact with the environment through sensors and 
actuators, typically in a feedback loop~\cite{AlurBook}. 
They use this feedback to adapt their behavior to the environment and achieve their goals. 
CPSs are used in a wide range of applications including smart homes, 
medical devices, and autonomous vehicles -- many of which
are safety-critical. 

CPSs, like software and hardware systems generally,
must operate according to specific functional 
requirements while also satisfying non-functional requirements. 
Engineers use various verification and validation processes 
such as reviews, testing, and formal methods to detect 
problems in such systems and to determine 
whether such systems meet their requirements~\cite{FisherBook}. 
In this work we focus on testing techniques. 

With CPSs, delaying testing activities until the final 
system is ready to operate in a real environment 
greatly increases the likelihood that serious 
faults will occur in that final system. 
Delaying testing also increases the possibility 
that such faults will be expensive, particularly
in safety-critical systems, due to the risks
to hardware components, equipment, and people 
that are involved in the final system. 
Early testing is imperative.

CPSs are often first specified iteratively 
as a sequence of model abstractions at 
different levels that can be tested 
using simulation models \cite{MatinnejadInpro2}.
Three common levels of abstraction (from most to 
least abstract) are: (1) Model-in-the-Loop (MiL), 
where the entire CPS is represented by a model;
(2) Software-in-the-Loop (SiL), where the model of
the controller is replaced with its code, but 
other components including hardware components
continue to be represented by models, and 
(3) Hardware-in-the-Loop (HiL), where the 
controller software is installed on the final 
platform and the other components, including
hardware components, are fully realized.
Testing CPS models at higher levels of
abstraction before continuing on to develop
the next level of models can result in earlier 
detection of faults and prevent those faults 
from propagating to the next levels, 
where they can be more expensive.

Many CPSs can be mathematically modeled by hybrid models~\cite{TahaBook}. 
One such model is a {\em hybrid automaton}, which is a 
formal model for a ``hybrid system'' -- a system with mixed 
discrete-continuous behaviour \cite{HenzingerBook}. 
Alur \cite{AlurInPro} defines ``hybrid systems''
as ``systems characterized by the interaction between 
discrete (digital) and continuous (physical) components.'' 
A hybrid model of a system can be designed 
using the dynamic equations for the system
as soon as these equations are known. 
Alur \cite{AlurInPro} notes that hybrid 
automatons ``are a central model for many CPS 
applications, from avionics to biomedical devices''.
For such CPSs, using a hybrid model as a basis
for testing requires no extra modeling effort
beyond that which is already required in the
construction of such systems.
(For additional information on hybrid models 
see Section \ref{sec:models}).

When test cases are generated from
hybrid models, they can target 
both continuous and discrete CPS behaviors 
\cite{ANTSAKLISInColl,HenzingerBook}. 
Further, it is difficult to predict all the 
conditions a CPS will encounter and 
test it under those conditions, due to the 
complicated interactions between the cyber 
and physical environments of a CPS \cite{SchneiderInPro}. 
The effect of those conditions, however --- whether external 
and environmental or internal --- is reflected 
in the values of the CPS's states, which are 
all captured in the hybrid model as invariants and 
guard conditions (this is also discussed further
in Section \ref{sec:models}.) 

Since hybrid models are designed at early stages of 
CPS development, i.e., at the system dynamics 
recognition stage, they allow test case generation 
to begin early in the design process of CPSs --
even before the simulation model of the CPS is designed. 

Several techniques for generating test cases for 
CPSs have been proposed; our earlier survey of
testing approaches for CPSs \cite{SadriArticle} discusses these.
Several authors have proposed techniques that use 
simulation models (e.g., \cite{AdeyemoInPro,MatinnejadArticle1, ArrietaInPro2,
MatinnejadInpro1, TurleaInPro, MatinnejadArticle, MatinnejadInpro2,
ZhaoArticle, LochauArticle, PohlheimArticle, ArrietaInPro3,
GadkariInPro, MenghiInPro, DeshmukhArticle}).
These techniques operate under the assumption that the 
simulation models have been designed correctly 
and that their correctness has been validated.
In this case the simulation models
can be used as ground truth models for 
generating test cases at the MiL level, 
and those test cases can then be used
to test the CPS at the SiL or HiL levels.  
In this case, however, testing {\em at the MiL level} 
using the generated test cases does not occur.

There are other techniques for generating test
cases for CPSs that generate test cases based 
on either the continuous or the discrete 
behavior of the CPS, but not both.
Badban et al. \cite{BadbanInPro} present a technique 
that targets the continuous behaviors of systems 
while Matinnejad et al. \cite{MatinnejadInpro2} present 
a technique that generates test cases for continuous controllers. 
These two techniques focus on continuous behaviors and do 
not target discrete behaviors when generating test cases. 
Bender et al. \cite{BenderArticle} propose a 
technique that generates test cases using discrete 
behaviors of the system and ignores its continuous behaviors. 
While such a technique may be fine in cases where
only one or the other type of behavior is present,
where both types are present the technique could miss important faults.

Several test case generation techniques for CPSs  
generate redundant test cases (e.g., \cite{MatinnejadArticle1,
ArrietaInPro2, MatinnejadArticle, MatinnejadInpro2, LochauArticle,
PohlheimArticle, ArrietaInPro3, GadkariInPro, MenghiInPro,
DeshmukhArticle, AlenaziInPro, GorenArticle, AraujoInPro,
CukicArticle, CarterInPro, MaArticle, KuroiwaArticle, SchneiderInPro, 
ZhangArticle, SinhaInPRo, ReniersInPro}); these incur unneeded costs.

A large number of test case generation techniques
(e.g., \cite{MurphyTech, MatinnejadInPro3, MarkiegiInPro2,LiArticle}) 
are non-CPS specific, are not guaranteed to work on CPSs, 
and have not been studied in CPS contexts.
More work -- including substantial empirical work,
is required to determine whether these techniques
might be applicable to CPSs or not. 
For additional information see our survey\cite{SadriArticle}. 

Some test case generation techniques operate under the
assumption that CPSs have a finite set of expected behaviors, 
and generate test cases based on this assumption. 
For example, Humeniuk et al. \cite{HumeniukInPro} 
expect the system under test to remain in 
each state for a predetermined amount of time. 
Such assumptions, however, do not typically hold
in practice, where unexpected conditions 
can cause a CPS to exhibit unexpected behavior. 
Some techniques (e.g., \cite{ChawlaInPro, MenghiInPro, 
RamezaniArticle, AnnapureddyInPro}) focus on generating 
test cases that are fault-revealing or falsifying, without 
attempting to assess whether the CPS functions as expected. 
When these techniques do not find failures,
we still cannot know whether the CPS behaves in accordance
with its requirements.
Finally, several techniques generate test cases for specific 
types of CPSs \cite{MoghadamInPro, KluckInPro, HumeniukInPro1,
FerdousInPro, StaraceInPro, CastellanoInPro, HumeniukInPro2, YanInPro}. 
As an example, Moghadam et al. \cite{MoghadamInPro} propose a technique 
that generates critical test roads on which to test an autonomous-driving car. 

When testing any software system it is important to have
a ``test oracle'' -- a method or device by which we can
determine whether test cases reveal faulty output or behavior.
Typical test oracles consider actual output or behaviors 
elicited by test cases and compare them to expected outputs or behaviors.
Where CPSs are concerned, it can be difficult or impossible
to codify what the expected output or behavior of the system is.  
CPSs are reactive systems that attempt to keep an ongoing 
interaction going in an acceptable way within an environment, 
rather than producing a final result upon termination \cite{HarelBook}. 
The standard correctness requirement for such systems
is that all executions must be allowed based on the 
system requirements and specifications \cite{CernInBook}, 
so a traditional test oracle that checks the correctness 
of the final results against a ground truth may issue 
a ``failure'' when the test results are not equal to 
the ground truth, whereas the CPS may have operated in 
an acceptable way and reached its goal.
For example, the goal of a robot could be to reach a
predefined target while avoiding obstacles in the environment, 
and in such a case we do not necessarily expect the 
robot to follow a specific and predefined trajectory. 
The expected behavior is simply to reach the target 
while avoiding the obstacles, irrespective of path.

Several CPS testing techniques (~\cite{TurleaInPro, 
ZhaoArticle, PohlheimArticle, GadkariInPro, AlenaziInPro, 
GorenArticle, CukicArticle, SchneiderInPro, ZhangArticle}) 
rely on manual test oracles or provide test oracles that 
issue verdicts by comparing test results against 
ground truths, which are usually expected outputs.
For reactive CPSs such oracles are problematic.

In this article we present {\sc HyTest}, a new technique for 
generating test cases for CPSs based on {\em hybrid models}.
{\sc HyTest} allows {\em pre-MiL level} test case generation, 
and the generated test cases can then be used to perform 
system testing of the simulation models of the CPS 
at the MiL and SiL levels or of the final product at the HiL 
level.\footnote{In this work, we focus primarily on the MiL level.}
As {\sc HyTest} generates test cases, it also employs 
an algorithm that reduces the incidence of redundant 
test cases, based on the hybrid model, rendering testing more efficient.
Finally, {\sc HyTest} provides a test oracle that uses 
the generated test cases to test the CPS as a reactive system. 
In other words, {\sc HyTest}'s test oracle checks 
whether the system operates correctly and in an acceptable 
manner, i.e., over an acceptable order of CPS states and transitions, 
to meet the goal or not. 

We assume that the hybrid model of the CPS under
test has been formally verified by test/control engineers
(i.e., checked for reachability and other problems \cite{HenzingerArticle}) 
prior to using {\sc HyTest} -- an assumption shared
by many other existing techniques. 
However, in some situations in which the hybrid model 
is not designed correctly, {\sc HyTest} can detect this,
display an error message, and end its execution. 

This work provides the following contributions:

\begin{itemize}
\item 
A novel test case generation technique, {\sc HyTest}, 
that generates and selects test cases for CPSs based on 
hybrid models and effectively and efficiently targets the mixed 
continuous and discrete behaviors of CPSs at the MiL level.
    
\item 
A novel test oracle that can automatically recognize
incorrect CPS behavior during simulation, using the 
states and transitions in the hybrid model, 
and issue an appropriate test verdict based on these. 

\item 
An empirical study examining the use of {\sc HyTest}
on simulation models developed in Simulink \cite{Simulink}
at the MiL level, that shows that {\sc HyTest}
was able to reveal more faults in a sample of CPSs as
efficiently (in a more complicated CPS) and more efficiently 
(in several less complicated CPSs) than a pair of baseline techniques. 
Our results show that overall {\sc HyTest} was able to
reveal all the faults in the CPSs considered in a reasonable amount of time.
Our results also show that our test oracle issued the correct 
test verdicts for all test cases, without any need for manual effort.
\end{itemize}

\section{Background}
\label{sec:background}

\subsection{Models}
\label{sec:models}
Models can express the structure and 
behavior of systems through conceptual 
or mathematical representations~\cite{scientificmodeling}. 
Models are useful for describing, developing, and 
validating systems such as CPSs. 
A {\em formal model} is a model that expresses the 
properties of a system at some level of abstraction~\cite{LimerkInPro}. 
Formal models typically represent systems using certain formalisms 
(e.g., linear temporal logic~\cite{emerson1990temporal}), and are created
prior to the development and deployment of such systems~\cite{NummenmaaArticle}.
A wide range of models have been utilized where software systems are concerned.
Here we describe two types of models that are relevant 
to CPSs and that we use in this work.

\subsubsection{Hybrid Models}

A hybrid automaton is a formal model that 
is used to represent a dynamical system with 
discrete and continuous components \cite{HenzingerBook}. 
A hybrid automaton is a labeled and directed graph 
(a finite state machine) that has the following 
components \cite{ANTSAKLISInColl,HenzingerBook}:

\begin{itemize}

\item  
$X$: A finite set $X = \{x_1,..., x_n\}$ of real-valued variables. 

\item 
$V$: A finite set of vertices, or discrete modes, indicating a control mode/location. 
All of these modes are ``acceptable'' because they are known/expected in the 
CPS and show the possible modes that a CPS can have while it is functioning.

\item 
$E$: A set of directed arcs or edges between vertices. 
An edge  $e \in E$ is also called a control switch or transition.

\item 
Flow condition: equations in the variables in $X$ describing continuous evolution of the system. 
While the hybrid automaton is in control mode $v\in V$, 
the variables $x_i$ change according to the flow condition.

\item 
Invariant condition: A condition under which the 
hybrid model may reside in control mode $v$.

\item 
Guard: An assignment of the variables in $X$. 
Each transition is associated with a guard. 
A transition is enabled when its assigned guard 
is true and its execution modifies the values 
of the variables according to the assignment. 

\item 
State: A state $\sigma = (v, x)$ of the hybrid automaton consists 
of a mode $v \in V$ and a continuous state $x_n\in X$. 
The state can change either by a discrete and instantaneous 
transition or over the passage of an interval of time through the continuous flow. 
A discrete transition changes both the control mode and the 
real-valued variables, while the passage of an interval of time changes only 
the values of the variables in $X$ according to the flow condition.

\end{itemize}

In a hybrid automaton model of a CPS, vertices model the 
discrete states, or modes, of the system while edges model its discrete dynamics or switches. 

Such an automaton can be used to design a controller 
and to develop a CPS, via simulation models or an implementation.

Figure \ref{fig:InvertedPendulumHybridModel}, taken 
(with small modifications) from
\cite{INHybrid}, depicts a hybrid model for a CPS.
(This model and the CPS to which it corresponds 
are described in Section \ref{subsec:CPS}.) 
In Figure \ref{fig:InvertedPendulumHybridModel}:

\begin{figure}[!t]
\centering
  \includegraphics[width=0.8\columnwidth]{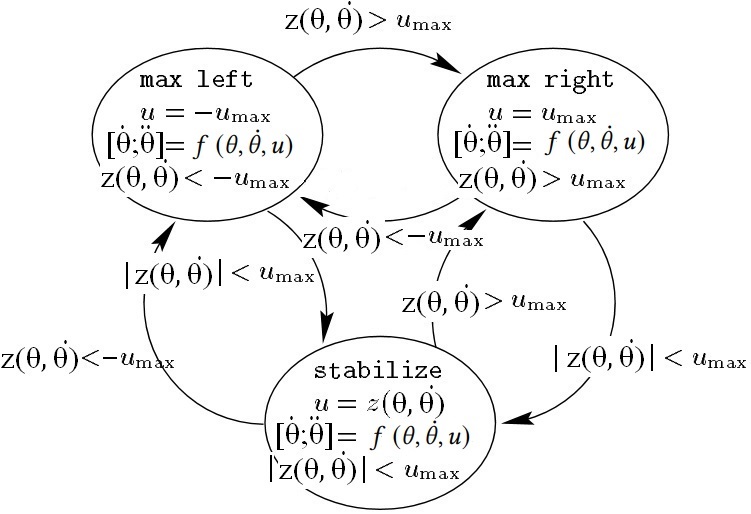}
  \caption{Hybrid model for an inverted pendulum \cite{INHybrid}.}
  \label{fig:InvertedPendulumHybridModel}
\vspace*{-12pt}
\end{figure}

\begin{itemize}
\item
$X=\{\theta, \dot{\theta}, u \}$ 

\item
$V=\{ \textit{``max \ left''}, \textit{``max \ right''}, \textit{``stabilize''}\}$ 

\item
$E=\{ \textit{``max left''} \rightarrow \textit{``max \ right''}, \textit{``max \ right''}\rightarrow \newline \textit{``max \ left''} \textit{``max \ left''}\rightarrow \textit{``stabilize''}, \textit{``max \ right''} \rightarrow \newline \textit{``stabilize''}, \textit{``stabilize''}\rightarrow \textit{``max \ left''}, \textit{``stabilize''}\rightarrow \textit{``max \ right''} \}$ 

\item
Flow conditions: $[\dot{\theta}; \ddot{\theta}]=f(\theta, \dot{\theta},u),  u= z(\theta, \dot{\theta}) \newline \text{ in} \textit{``stabilize''},  
u=-u_{max} \text{ in} \textit{``max left''} ,u=u_{max}\text{ in}  \newline \textit{``max right''}$ (See [12] for more details).

\item
Invariant conditions: $\{ z(\theta, \dot{\theta})<-u_{max}, z(\theta, \dot{\theta})>u_{max}, \newline |z(\theta, \dot{\theta})|<u_{max} \}$

\item
Guards: $\{ z(\theta, \dot{\theta})<-u_{max}, z(\theta, \dot{\theta})>u_{max}, |z(\theta, \dot{\theta})|<u_{max} \}$. 

\end{itemize}

\noindent

\subsubsection{Simulation Models}

Simulation models typically form the 
basis of computer simulations~\cite{DuranArticle}. 
Simulation models can be executed in a virtual 
environment to demonstrate the behavior of a
system before that system has been implemented. 
Tests can be performed as simulations
on simulation models~\cite{Simulation}.

\vspace*{-6pt}
\subsection{Model-Based Testing}

{\em Model-based testing} is typically
performed prior to system development and deployment 
allowing test engineers to examine whether a system's model 
conforms to the system's specifications. 
Model-based testing techniques typically generate test cases 
from structural or behavioral models of a system.~\cite{BroyBook}. 
Such {\em model-based test cases} are typically
abstract, and require additional detail relevant
to a system's implementation to be
added to them; this transforms 
them into {\em concrete test cases} that can
be executed on an implemented system 
or its more detailed models. 
Model-based test cases are typically easier to 
maintain than code-based test cases, and 
they can be used to measure the 
{\em coverage} (of the model) achieved in testing.

\vspace*{-6pt}
\subsection{Extended Example} 
\label{subsec:CPS}

\begin{figure}[!t]
\vspace*{-12pt}
\centering
  \includegraphics[width=0.6\columnwidth]{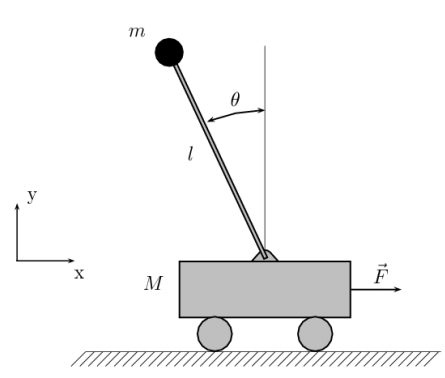}
  \caption{Inverted Pendulum \cite{OlfaInPro}}
  \label{fig:invertedPendulum}
\vspace*{-12pt}
\end{figure}

To illustrate {\sc HyTest} in this article, we utilize the 
simple example presented in \cite{OlfaInPro,INHybrid,SadriArticle}. 
Figure \ref{fig:invertedPendulum} 
depicts a system that includes an inverted pendulum of length $l$ 
and mass $m$ mounted on a cart of mass $M$. 
A force, $F$, is applied to the cart and drags 
it forward or backward to balance the pendulum. 
To maintain stability (i.e., a balanced pendulum), 
a control input is computed and sent to the motors 
in the cart's wheels periodically based on 
feedback the sensors provide, i.e. information about 
the angle, $\theta$, and angular velocity, 
$\dot{\theta}$, of the pendulum.

A general template for a test case for the inverted pendulum
system can be written as follows:
\[
testcase_{i,j}= \left\{ F_i, \theta_i, \dot{\theta}_i, x_i, \dot{x}_i, \theta_j, \dot{\theta}_j, x_j, \dot{x}_j \right\}.
\]

\noindent
In a system such as this, a template 
test case at sampling time, $i$, consists of a 
set of one or more test inputs, (possibly
empty) sets of pre-condition(s), expected 
output(s), and expected post-condition(s). 
In this system, force $F_i$ is the 
test input, while the angle $\theta_i$ and angular 
velocity $\dot{\theta}_i$ of the pendulum, as well 
as the position $x_i$ and velocity $\dot{x}_i$ 
of the cart along the $x$-axis, are pre-conditions. 
The expected test outputs are $\theta_j$ ($j$ can 
be the same as $i$) and $x_j$, and the expected 
post-conditions are $\dot{\theta}_j$ and $\dot{x}_j$. 
Other properties such as the state of the system, 
its failure or success, or safety properties such 
as its stability at a given time $j$, 
could also be expected post-conditions -- we omit
these for simplicity.
A specific test case replaces the variables in the template with concrete values.

The hybrid model for this example, shown in 
Figure \ref{fig:InvertedPendulumHybridModel}, 
has three acceptable modes (shown as ellipses) 
that model its discrete states, and six 
edges/transitions between modes that model 
its discrete dynamics or switches.
Inside each mode, the first line displays the mode's name.  
The next two lines display flow conditions, 
and the fourth displays an invariant condition. 
The labels on each edge display guard conditions. 
The function $z(\theta, \dot{\theta})$, which calculates the total 
energy of the system, together with $u_{max}$, 
which is a constant and shows the maximum total 
energy that the CPS can have in order to remain 
stable, determine whether the CPS should remain 
in the same mode or change to another one. 
In other words, if the absolute value of $z(\theta, \dot{\theta})$ 
is in the range of $-u_{max}$ to $u_{max}$, then 
the inverted pendulum is stable; otherwise it 
is falling down, either left or right.

This hybrid model describes how
the inverted pendulum functions, as follows.
The CPS can begin in any of the acceptable modes. 
Here we assume that the pendulum is initially in the upright 
position, i.e., mode \textit{stabilize}, and when released
the pendulum moves towards the left. 
Suppose force $F$ moves the cart to the right to balance 
the total energy of the CPS and keep the pendulum 
stable, captured by the mode \textit{stabilize}; 
in this case, the pendulum moves to the right. 
Now force $F$ moves the cart to the left to balance 
the total energy of the CPS and stabilize the pendulum. 
If, for some reason, the total energy of the CPS is 
less than $-u_{max}$, i.e., the pendulum is falling 
to the left, the system will be in mode \textit{max left}. 
Depending on the acceptable deviation the controller 
designer has considered for the modes \textit{max left} and 
\textit{max right}, the controller may or may not be able 
to stabilize the pendulum and return the CPS to mode \textit{stabilize}.

\section{Approach}
\label{sec:approach}

\noindent Algorithm \ref{alg:approachOverview} 
provides an overview of {\sc HyTest}. 
{\sc HyTest} receives information about a CPS's hybrid 
model, conditions that indicate the CPS's failure, 
simulation parameters, information on the CPS's dynamics, 
and a simulation model of the CPS.
All of these are 
either required to design and develop the 
CPS or can be obtained easily from its requirements. 
The algorithm outputs the faults that are 
revealed during the CPS testing process.

\SetKwComment{Comment}{/* }{ */}
\SetKwInput{KwData}{Inputs}
\SetKwInput{KwResult}{Output}

\begin{algorithm}[t]
\small
\caption{{\sc HyTest}}\label{alg:approachOverview}
\KwData{hModel: Hybrid Model, \newline
unConditions: Unacceptable Conditions, \newline
simData: Simulation Data, \newline
sysDyn: System and CPS Dynamics, \newline
simulationModel: Simulation model}
\KwResult{faults: Detected faults in the CPS}

hModel=getHybridModelData();\label{alg:11}

unConditions= getUnacceptableConditions();\label{alg:12}

simData= getSimulationData();\label{alg:13}

sysDyn= getDynamics();\label{alg:14}

cGraph= genConditionsGraph(hModel,unConditions);\label{alg:15}

tConditions= genTestConditions(cGraph);\label{alg:16}

testcases=genTestCases(tConditions,simData,sysDyn);\label{alg:17}

[faults]=testCPS(testcases,simData,tConditions,simulationModels);\label{alg:18}

\KwRet{$faults$}

\end{algorithm}

{\sc HyTest} begins (line~1) by retrieving 
data about the hybrid model of the CPS.
This data includes the total number of modes, 
the CPS goal(s), invariant conditions, guard 
conditions that show the transitions between modes, 
and variables in the hybrid model including 
their acceptable value range and precision. 
Next (line~\ref{alg:12}), {\sc HyTest} obtains a list 
of ``unacceptable conditions'': these are conditions 
that are not expected to occur during the operation 
of the CPS, or that would lead to a failure. 
If no unacceptable conditions are provided for the CPS, 
{\sc HyTest} temporarily sets the variable {\em unConditions} to null
(Section \ref{sec:unacceptable} describes this step further). 
{\sc HyTest} next (lines~\ref{alg:13}-\ref{alg:14}) retrieves 
data about the simulation model used to test 
the CPS, and about the system's and CPS's dynamics
including its state-space representation; transfer 
functions or Matlab code that implements the 
system's dynamics; and initial values, inputs, 
and simulation time, provided as Matlab ``init'' files.
Using this information, {\sc HyTest} creates 
a ``condition graph'' (line~\ref{alg:15})
and using this graph, generates 
test conditions (line~\ref{alg:16}). 

{\sc HyTest} uses test conditions to partition the 
input space and then generate test cases (line~\ref{alg:17}). 
It passes these test cases, along with simulation data,
to the testCPS function (line~\ref{alg:18}).
The testCPS function implements a testing framework 
with which to test the CPS by running its simulation 
model, and ultimately returns the faults that are 
revealed by test cases during testing.  
We explain each of the foregoing steps 
in the subsections that follow.  

\vspace*{-6pt}
\subsection{Unacceptable Condition Determination}
\label{sec:unacceptable}

To generate test cases and provide a test oracle,
{\sc HyTest} utilizes ``acceptable'' and ``unacceptable'' 
conditions on variables in the CPS -- we assume acceptable and, 
possibly, unacceptable conditions have all been identified by 
the system/controller designer and are available to {\sc HyTest}.
{\em Acceptable conditions} are those that are 
known in the target CPS, lead to expected behavior 
of the CPS, and do not cause the CPS to fail. 
These conditions include all the invariants and guards found in 
the CPS hybrid model -- in our example $|z(\theta, \dot{\theta})| <u_{max}$. 
{\em Unacceptable conditions} are those conditions that are not 
known/expected in the target CPS or that lead the CPS to a failure. 
In our example, these include range values for the variables 
in the CPS, e.g., the displacement of the cart from its initial position is greater 
than \unit[3]{m}, so $|x| > 3$ is an unacceptable condition. 
In Section \ref{sec:condition} we explain what {\sc HyTest} 
does when unacceptable conditions are not known. 

As noted earlier, it is difficult to predict all 
of the conditions a CPS will encounter, but the effect 
of those conditions is reflected in the values of 
the CPS's states, which are all captured in the 
hybrid model as invariants and guard conditions. 
Because conditions, whether acceptable or unacceptable, 
set limitations and bounds on variables in the CPS, 
and because the fact that a condition is acceptable or not 
determines whether the CPS functions correctly or not, 
we believe that generating test cases using conditions 
will be effective for revealing faults in CPSs.

\vspace*{-6pt}
\subsection{Condition Graph Generation}
\label{sec:condition}

To generate test cases using information about 
the hybrid model and unacceptable conditions, 
{\sc HyTest} generates a specific type of 
graph that we call a {\em condition graph}. 
A condition graph is a representation of a CPS 
that shows all acceptable, final, and failing modes 
in the CPS, along with all conditions that change 
the mode of the CPS from one to another and 
may lead the CPS to a failure or success.

\SetKwComment{Comment}{/* }{ */}

\begin{algorithm}[b]
\small
\caption{Condition graph generation}\label{alg:conditionsGraphGen}

  \SetKwInOut{Input}{inputs}
  \SetKwInOut{Output}{output}
  \SetKwProg{genConditionGraph}{genConditionGraph}{}{}

  \genConditionGraph{$(hModel,unConditions)$}{
    \Input{hModel: Hybrid Model, \newline
    unConditions: Unacceptable Conditions}
    \Output{cGraph: Conditions Graph}
    
    cGraph=addModes($hModel$);\label{alg:211}
    
    cGraph=determineCPSGoal($hModel,cGraph$);\label{alg:212}
    
    cGraph=addFailingMode($cGraph$);\label{alg:213}
    
    cGraph=addInvariants($cGraph, hModel$);\label{alg:214}
    
   cGraph=addUnacceptableConditions($cGraph, hModel$);\label{alg:215}

   cGraph=addEdges($cGraph, hModel$);\label{alg:216} 
    
    cGraph=addLabels($cGraph, hModel$);\label{alg:217}
    
    \KwRet{$cGraph$}\;
  }
\end{algorithm}

{\sc HyTest} generates a condition graph in two 
sub-steps using the module genConditionGraph, 
shown in Algorithm \ref{alg:conditionsGraphGen}. 
genConditionGraph receives a hybrid model and 
set of unacceptable conditions as its inputs 
and returns the condition graph. 
If unacceptable conditions are not known, the 
conjunction of the negation of all acceptable 
conditions are considered to be unacceptable conditions. 
For example, if the unacceptable condition is not known 
in our inverted pendulum example and $|x| <= 3$ and $x \sim = 0$ 
are two acceptable conditions, then {\sc HyTest} 
considers $ \sim (|x| <= 3)\ \&\& \sim (x \sim= 0) $ 
to be an unacceptable condition.

\begin{figure}[t]
\centering
  \includegraphics[width=.95\columnwidth]{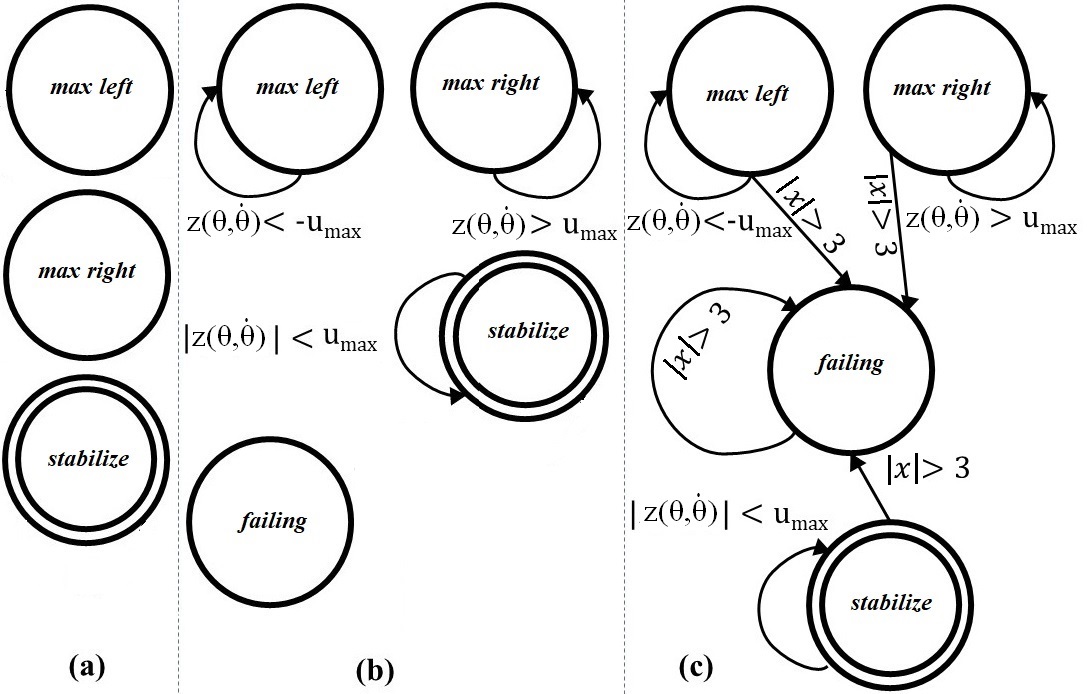}
  \caption{Condition Graph Generation, Step 1: {\sc HyTest} (a)  adds modes to the graph and recognizes the final mode(s);  (b) adds one more mode to the graph as the failing mode and adds invariant conditions of the modes as self-loops; (c) adds unacceptable conditions to the edge between the acceptable and failing modes and to the self-loop on the failing mode.}
  \label{fig:substep1abc}
\vspace*{-12pt}
\end{figure}

In its first step, for each mode in the hybrid model, 
{\sc HyTest} adds the mode to the graph and adds 
its invariant conditions as self-loop edges (line~\ref{alg:211}). 
Next, if the goal conditions are the invariant(s) of the mode(s), 
{\sc HyTest} designates one or more acceptable 
modes as ``final modes'' (line~\ref{alg:212}) based on 
the conditions that the CPS designer sets as the goal of the CPS; 
otherwise it just saves the goal conditions for the next steps.
In our inverted pendulum example, the goal of the final CPS 
is to stabilize the pendulum in the upward position and 
keep it in this position, so the condition of the final 
mode is $|z(\theta, \dot{\theta})|<u_{max}$ and {\sc HyTest} designates the 
mode ``stabilize'' as the final mode. 
The result of this step is
shown in Figure~\ref{fig:substep1abc}(a). 

Next (line~\ref{alg:213}), {\sc HyTest} adds 
one extra mode to the set of CPS 
modes, designated as a ``failing'' mode.
For any mode in the graph, {\sc HyTest} adds its 
invariant condition as a self-loop (line~\ref{alg:214}). 
The result of this step is shown 
in Figure~\ref{fig:substep1abc}(b). 

Next (line~\ref{alg:215}), {\sc HyTest} 
generates the complete unacceptable conditions expression and augments 
the graph as follows. First, using the logical operator ``|'', {\sc HyTest} 
joins all unacceptable conditions in the CPS and considers 
them as initial unacceptable conditions expressions. 
For our example, the initial unacceptable condition is $ (|x| > 3)$.
Second, for each mode that has unacceptable conditions, 
using the logical operator ``|'', {\sc  HyTest} combines 
unacceptable conditions for that mode with the initial expression. 
Our example did not include any unacceptable conditions of this type, 
but if the unacceptable condition of this type was 
$ (x == 0) $ then the 
unacceptable conditions expression would be  
$ (|x| > 3) \ || \ (x == 0)$. 
{\sc HyTest} adds this unacceptable 
condition expression to the graph on new edges from acceptable 
modes to failing modes and on a self-loop edge on the failing mode. 
In our Inverted Pendulum example, the unacceptable 
condition expression is $|x| > 3$, shown in 
Figure~\ref{fig:substep1abc}(c) as variable $d$. 
If there are no unacceptable conditions recognized 
by the system/controller designer or generated 
by {\sc HyTest}, then there is no unacceptable conditions expression and the failing mode will be a separate graph 
that includes just one node, and will be ignored by {\sc HyTest}. 
In such a case {\sc HyTest} recognizes faults in the CPS 
(if there are any) and issues ``failed'' verdicts as 
explained in Section \ref{sec:oracle}. 

\begin{figure}[t]
\centering
  \includegraphics[width=.9\columnwidth]{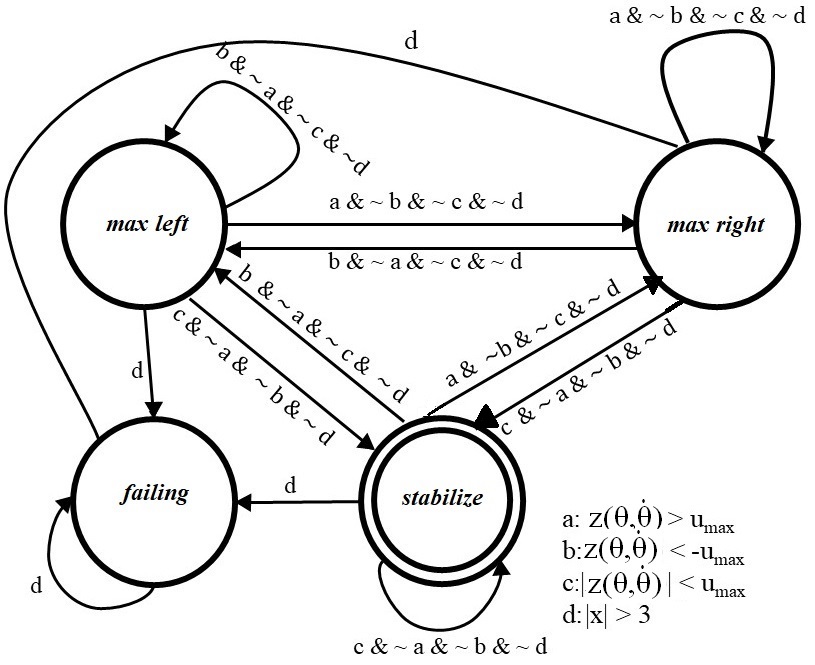}
  \caption{Condition Graph Generation, Step 2}
  \label{fig:substep2}
\vspace*{-6pt}
\end{figure}

In its next step (line~\ref{alg:216}), {\sc HyTest} 
adds all the edges that exist between modes 
in the hybrid model that are missing in the 
condition graph, with no labels assigned to them. 
For example, by comparing Figure \ref{fig:substep1abc}(c) 
with the hybrid model, i.e. Figure \ref{fig:InvertedPendulumHybridModel}, 
we see that  the edge from mode \textit{max left} to mode 
\textit{max right} in the hybrid model is missing 
and must be added to the condition graph.
Next (line~\ref{alg:217}), the algorithm 
adds labels or completes them as follows. 
First, if an edge has no label, the algorithm
adds its guard conditions as its label. 
Second, for all edges other than those that end in failing 
mode, using the logical operator ``\&'', the algorithm 
joins the label of the edge to the negation of the 
labels of all other outgoing edges of the mode. 
Then it adds the negation of unacceptable conditions expression, 
which is generated in line (line~\ref{alg:215}), 
using the same logical operator, i.e. ``\&''.
For example, the label added to the edge from mode 
\textit{max left} to mode \textit{max right} 
is $z(\theta, \dot{\theta})\ >\ u_{max}$, and 
the algorithm adds $ \& \sim z(\theta, \dot{\theta})\ 
<\ -u_{max}\ \& \sim |z(\theta, \dot{\theta})|\ < 
\ u_{max} \& \sim |x| \ > \ 3$ to that edge.
Figure \ref{fig:substep2} shows the condition graph 
after this step -- for readability, complex 
conditions have been represented by variables
as shown in the legend within the figure.

{\sc HyTest} has now generated the complete 
condition graph, as shown in Figure \ref{fig:substep2}. 
This graph is the input for the next step. 

\vspace*{-6pt}
\subsection{Test Condition Generation}

\emph{Test conditions} are the conditions in the 
condition graph that are used to generate test 
inputs and to determine whether the generated 
test inputs lead the CPS into final, acceptable, 
or failing modes.

\begin{algorithm}[t]
\small
\caption{Test condition generation}\label{alg:testConditionGen}

  \SetKwInOut{Input}{inputs}
  \SetKwInOut{Output}{output}
  \SetKwProg{genTestConditions}{genTestConditions}{}{}

  \genTestConditions{$(cGraph)$}{
    \Input{cGraph: Condition Graph}
    \Output{tConditions: Test Conditions}
    tConditions=[];
    
    \ForEach{edge $e_i \in cGraph$}{\label{alg:21}
      $temp = source + ``,`` +destination + ``\#``+label_i$\;\label{alg:22}
      
      \uIf{$ type(destination) == failing$}{\label{alg:24}
        $temp.append(``@failed``)$\;
      }
      \uElseIf{$ type(destination) == final$}{
        $temp .append(``@passed``)$\;
      }\uElseIf{$ type(destination) == acceptable \ \&\& 
      \newline $\texttildelow$ isEmpty(cGraph.finalMode)$}{ \label{alg:27}
      $temp.append(``@acceptable``)$;
      }
      \Else{
      $temp1.append(``\&$\texttildelow$(``,goalCondition,``)...
      \newline ...@acceptable``$;
      
      $tConditions.add(temp1);$
      
        $temp.append(``\&(``,goalCondition  ,``)@passed``)$;

      }\label{alg:25}
      $tConditions.add(temp);$\label{alg:26}
    }
    \KwRet{$tConditions$}\;
  }

\end{algorithm}

Algorithm \ref{alg:testConditionGen} presents our algorithm, 
genTestConditions, for generating test conditions. 
genTestConditions takes a condition graph as input. 
For each edge in the graph (line~\ref{alg:21}), 
genTestConditions obtains its source and destination modes 
and concatenates them as a formatted string 
(line~\ref{alg:22}), which is easy to parse
when the approach needs to extract specific 
data from the test conditions. 
Next, genTestConditions obtains the edge's label, which 
shows under what conditions the CPS changes 
mode from source to destination. 
These conditions are used later to generate 
values for the CPS variables to make the 
system change from one mode to another, so 
the algorithm adds the condition to the 
formatted string (line~\ref{alg:22}). 
Recall that {\sc HyTest} categorizes the 
modes as acceptable (any mode that is neither 
final nor failing), final, or failing. 
Using these categories, the algorithm recognizes 
(lines~\ref{alg:24}-\ref{alg:25}) the type of the 
destination mode (of the edge) and appends it to the 
formatted string (line~\ref{alg:26}), which helps the 
test oracle issue the correct test verdict for the test inputs. 
Recall also that if the goal conditions of the CPS 
are not the invariant(s) of any mode, then there 
is no final mode and {\sc HyTest} saves the goal condition(s). 
In this situation, the condition graph has just two 
types of modes: failing and acceptable. 
If the CPS is in any of the acceptable modes and 
the values of the CPS's variables fit the goal 
condition(s) then the system has reached the goal 
and the state (which in the CPS includes mode and 
values) is considered to be final. 
Otherwise, it is considered to be acceptable. 
This is how lines ~\ref{alg:27}-\ref{alg:25} 
identify whether a test condition is an acceptable or passing one.

The following line illustrates a sample test 
condition that is generated by genTestConditions for our example 
and corresponds to the self-loop on the mode \textit{stabilize}.

\vspace*{6pt}
\noindent
$\textit{stabilize},\textit{stabilize} \ \# \ (|z(\theta, \dot{\theta})| < u_{max}) \ \& \sim(z(\theta, \dot{\theta}) > u_{max}) \ \& \sim(z(\theta, \dot{\theta}) < -u_{max} ) \ \& \sim(x > 3)@passed$

\vspace*{-6pt}
\subsection{Test Case Generation}
\label{sec:TCGen}

Algorithm \ref{alg:TCGenOverview}, genTestCases, generates test cases.
The algorithm receives test conditions, simulation data, 
and the CPS's and system's dynamics as input data, and outputs generated test cases.

\begin{algorithm}[t!]
\small
\caption{Test case generation}\label{alg:TCGenOverview}

  \SetKwInOut{Input}{inputs}
  \SetKwInOut{Output}{output}
  \SetKwProg{genTestCases}{genTestCases}{}{}

  \genTestCases{$()$}{
    \Input{tConditions: Test Conditions,\newline
    simData: Simulation Data,\newline
    sysDyn: CPS's and system's Dynamics}
    \Output{testcases: Generated Test Cases}

[TC1,testConditionsCovered]=genTestcasesCPS(\newline testConditions,simulationData,CPSDynamics);\label{alg:31}

[TC2,testConditionsCovered]=genTestcasesSystem(\newline testConditions,simulationData,systemDynamics,\newline testConditionsCovered);\label{alg:32}

TC3=genTestcasesUncovTestConditions(\newline testConditions,testConditionsCovered);\label{alg:33}

testcases= select(TC1,TC2,TC3); \label{alg:34}

\KwRet{$testcases$}\;
}
\end{algorithm}

The genTestCases algorithm proceeds in four major steps. 
In Step~1 (line~\ref{alg:31}), genTestCases 
invokes algorithm genTestCasesCPS. 
genTestCasesCPS (Algorithm \ref{alg:TCGen1})
uses test conditions, simulation data, 
and the CPS's dynamics to generate an initial set of test cases. 
To this end, genTestCasesCPS (line~\ref{alg:311})
first obtains the response of the CPS to the control 
input provided as part of the CPS's dynamics in 
its init file, and obtains the outputs. 
The output values are either values of 
the variables in the hybrid model, such as  
$\theta$, or values that are used to calculate 
the values of the expressions in the 
hybrid model, such as $\cos{\theta}$. 
The algorithm checks all of the test conditions to 
find the outputs that fit them (line~\ref{alg:312}). 
(In other words, it uses the test conditions to partition the input space.)
If the algorithm finds any output that fits a given
test condition $t_i$, then it marks that test condition 
as covered (line~\ref{alg:313}) and extracts the 
type of the destination mode of the test condition, 
by extracting anything after the sign ``@'' from 
the test condition $t_i$, (line~\ref{alg:314}). 
This is considered to be the type of the initial mode 
of the CPS, and is used later by the test 
oracle to issue correct test verdicts. 
For each output that fits test condition $t_i$, the 
approach considers the output values as test inputs 
and generates a set $(outputs,type$ $of$ $initial$ $mode,t_i)$ 
as a test case and appends it to the test 
suite (lines~\ref{alg:315}-\ref{alg:316}). 
If there is at least one output that does not fit 
any test conditions, this means that the hybrid model 
provided is incorrect (line~\ref{alg:317}).

\SetKwComment{Comment}{/* }{ */}

\begin{algorithm}[t]
\small
\caption{Test case generation, Step 1}\label{alg:TCGen1}

\SetKwInOut{Input}{inputs}
  \SetKwInOut{Output}{output}
  \SetKwProg{genTestcasesCPS}{genTestcasesCPS}{}{}

  \genTestcasesCPS{$(testConditions,simulationData,\newline CPSDynamics)$}{
    \Input{Test Conditions,Simulation Data,CPS's Dynamics}
    \Output{testCases1: Test Cases Generated using the Controlled Environment,\newline testConditionsCovered: Test Conditions that are Covered by test suite testCases1}

    cpsOutputs=getCPSResponse(CPSDynamics);\label{alg:311}
    
    \ForEach{testCondition $t_i \in testConditions$}{
    
      indices=\FuncSty{findOutputs($t_i,cpsOutputs$);}\label{alg:312}
      
      \uIf{\texttildelow isEmpty(indices)}{
      
        $testConditionsCovered_{i}$= true; \label{alg:313}
        
        initialMode=\FuncSty{extractAfter($t_i,@$);} \label{alg:314}
        
          \ForEach{index $j \in indices$}{\label{alg:315}
          
          $newTC =(cpsOutputs_j, initialMode,i)$;
          
          testCases1= \FuncSty{append($testCases1,newTC$);}
          }\label{alg:316}
        }
      }
     
   \uIf{there is any output that does not fit any test conditions}{
      print(``Wrong hybrid model!!!'');\label{alg:317}
      }
\KwRet{$testCases1$}\;
}
\end{algorithm}

In Step~2 (line~\ref{alg:32}), genTestCases uses 
the uncontrolled system's dynamics along with the 
other inputs to generate a second set of test cases. 
The algorithm genTestcasesSystem() generates test cases in 
a manner similar to that used by genTestcasesCPS(). 
There are two differences between these algorithms. 
First, genTestCasesSystem generates test cases using 
the dynamics of the uncontrolled system, whereas 
genTestcasesCPS() uses the dynamics of the controlled system.
Second, to avoid generating redundant test 
cases, genTestCasesSystem checks test conditions 
that are not covered by the test cases that 
have been generated by genTestcasesCPS(). 

In Step~3, if there are any test conditions that are not covered 
by the test cases generated in Step~1, these are
used to  generate additional test cases (line~\ref{alg:33}), 
i.e., a third set of test cases.
In the first and second steps, all test conditions 
that are covered by outputs are marked, and it is possible
to have test conditions that are not covered by any outputs. 
To cover these uncovered test conditions, {\sc HyTest} first 
uses the variables in the hybrid model, their acceptable 
value ranges, and their precisions to generate an input 
space that includes valid and invalid values 
for the variables in the hybrid model. 
For each variable, {\sc HyTest} takes the acceptable 
ranges of values, in the form of the range's boundaries,
and generates another range in two steps:

    $ 1) maxBound=max(|boundries_i|)$

    $ 2) -2*maxBound:precision:2*maxBound $

\noindent
This ensures that the resulting range includes 
invalid values as well as valid ones. 
In the second step, we can replace 2 with any 
number greater than 2, but because the goal of 
this step is to provide a range of valid and 
invalid values and $maxbound$ and $minbound$ 
are selected based on the boundaries of acceptable 
values, using a co-efficient greater than 
2 just increases the range of invalid values. 
Since the test conditions are used to partition the
input space, increasing the range of invalid values 
only results in larger numbers of invalid inputs 
without affecting the fault revealing ability of 
{\sc HyTest} (as we explain later in this article). 
This unnecessarily increases the time required
to test the target CPS. 

Next, {\sc HyTest} calculates the Cartesian product of the variables' 
value sets to find all possible combinations. 
Finally, it finds the test conditions that each resulting 
combination may fit in and generates test cases in
the same manner as is done for the first
and second classes of generated test cases.

Finally, in Step~4, because the first three steps of genTestCases may return 
redundant (identical) test cases, genTestCases attempts to remove 
redundant test cases through a selection process. 
Before this process begins, genTestCases has 
classified the generated test cases using test conditions. 
Given this classification, the selection process proceeds in two steps. 
First, the process selects one test case per class (i.e., per test 
condition) in order to cover all the test conditions, 
i.e., all partitions of input spaces,
which are all of the transitions in the condition graph. 
The test suite resulting from this step, however, may still 
contain redundant test cases depending on the extent to
which invariants and guard conditions overlap.
This necessitates a second step, in which genTestCases removes 
the redundant test cases from the test suite by finding the 
unique ones in terms of test inputs and initial modes. 

A concrete test case for our inverted pendulum example looks like this:

$testcase_i=$ 
$\{-0.495502602162482, 2.88465092291609, $ \newline \indent
$ 0.529831722628956, 1.33637906750083, passed \}$.

\noindent
The members of this set are the pendulum's angle, 
the pendulum's angular velocity, the cart position, 
the cart velocity, and the expected type of the mode 
that the CPS will be in when the its variables are 
set to these values (in this case the mode is passed ). 

\vspace*{-6pt}
\subsection{Test Execution and Test Oracle Operation}
\label{sec:oracle}

The final step of Algorithm~\ref{alg:approachOverview} (line~\ref{alg:18}) invokes algorithm testCPS, which executes generated test cases on the CPS.
Algorithm \ref{alg:testFramework} presents our algorithm, testCPS, that 
provides details on this step of the process. 

\SetKwComment{Comment}{/* }{ */}

\begin{algorithm}[t!]
\small
\caption{Test execution and test oracle}\label{alg:testFramework}

  \SetKwInOut{Input}{inputs}
  \SetKwInOut{Output}{output}
  \SetKwProg{testCPS}{testCPS}{}{}

  \testCPS{$(testcases,simData,)$}{
    \Input{testcases: Test Cases,
    \newline simData: Simulation Data,
    \newline tConditions: Test Conditions,
    \newline simulationModels: Simulation Models of CPS}
    
    \SetKwProg{try}{try}{:}{}
    \SetKwProg{catch}{catch}{:}{end}
    \Output{faults: Revealed Faults}
    
    testInputs=extractTestInputs($testcases$); \label{alg:51}
    
    initialMode= extractInitialMode($testcases$); \label{alg:52}
    
    \ForEach{test input $input_i \in testInputs$}{\label{alg:530}
    
    \try{}{
    
      testOutput=simulateModel($input_i,simData$); \label{alg:53}
      
      possibleCPSModes=recognizeCPSModes(...
      ...$testOutput,tConditions$); \label{alg:54}
      
      \ForEach{mode $mode_j \in possibleCPSModes$}{\label{alg:550}
      
          \uIf{$mode_j == null$}{\label{alg:55}
          
              print(``The hybrid model was not designed correctly.'');\label{alg:551}
          }
          \uElse{
            \uIf{\texttildelow isAllowed($mode_j$, $mode_{j+1}$)}{\label{alg:56}
                $faults_i \gets true$ \label{alg:561}
            }
            \uElseIf{$mode_j$ contains failing}{\label{alg:57}
            
                $faults_i \gets true$ \label{alg:571}
                
            }
         }
        }
        \uIf{$possibleCPSModes$ \texttildelow contains ``failing''}{\label{alg:590}
        
            $lastMode \gets the last mode\ in \ possibleCPSModes$;
        
            \uElseIf{ $lastMode$ contains final}{
            
                \uIf{$initialMode ==  final \ ||\  initialMode == acceptable$}{
                
                    $faults_i \gets false$ \label{alg:59}
                }
                \uElse{
                    $faults_i \gets true$ \label{alg:60}
                    
                }
            }
            \uElse{
                $faults_i \gets true$ \label{alg:62}
                
            }
       }   
      }\catch{Exception}{\label{alg:63}
        
        $faults_i \gets true$\label{alg:631}
      }
      
    }
    \KwRet{$faults$}\;
  }

\end{algorithm}

The testCPS algorithm takes the generated test 
suite and simulation data as inputs and returns 
faults revealed during testing.
testCPS begins by retrieving test cases and 
extracting test inputs (line \ref{alg:51}) 
and the initial modes (line \ref{alg:52}). 

As already noted, the hybrid model has already 
been verified; i.e., all its modes/states are reachable, 
so {\sc HyTest} puts the CPS in different modes 
and transitions without any concern about their reachability. 
testCPS, on lines \ref{alg:530}-\ref{alg:53}, for each test case, 
initially sets the values of the variables to the test 
input values of test cases to put the CPS in 
every possible acceptable, final, or failing 
mode and then checks the behavior of 
the CPS until the end of simulation. 

As an example, using the concrete test case for our inverted pendulum CPS 
(which we have presented in Section \ref{sec:TCGen}), {\sc HyTest} starts simulating the CPS from a ``passing'', 
i.e., final, state by setting the initial values of the 
CPS's variables to the values in the test case and 
simulates the CPS for the entire simulation time. 

For each test input, testCPS receives the simulation 
output(s) as a set(s) of values with the length of the 
(simulation time)/(sampling time). 
Then, testCPS inspects the simulation outputs 
to see whether the CPS has stopped at a final mode 
through sequences of allowed transitions and whether
the values of the variables fit the goal conditions. 

To issue a test verdict for test 
cases, the test oracle behaves as follows. 
First, using the test conditions and the test output 
signal, the test oracle recognizes the possible modes 
of the CPS at any sampling time (line \ref{alg:54})
by checking the test outputs against test conditions. 
Since a test output at time $j$ may fit in several test 
conditions, the CPS at time $j$ can be in different modes 
and the real mode of the CPS depends on its real condition at testing time. 
At this point (lines \ref{alg:550}-\ref{alg:55}), testCPS checks whether 
there is any test output with no modes assigned to it. 
In this case, the hybrid model was designed incorrectly, line\ref{alg:551},
because there is a behavior in the CPS that does not 
fit any modes/transitions represented in the hybrid 
model, i.e., the hybrid model missed a mode/transition. 
According to the modes in the condition graph, the 
modes and transitions (based on their destinations) of the CPS can be categorized into three 
groups: acceptable, final, and failing. 
Based on these groups, the test oracle issues 
test verdicts for the test cases as follows:

\begin{itemize}

\item 
Using the condition graph, the test oracle determines 
(lines \ref{alg:56}-\ref{alg:561})
whether transitions between consecutive modes are 
allowed or not, i.e., transitions between 
modes of the test output at times $i$ and $i+1$. 
If the transition is not allowed then a failure 
occurred, e.g., there is an overshoot in the CPS output values.
    
\item 
If the possible modes of the CPS include ``failing'' 
in at least one sampling time, the test oracle issues 
a ``failed'' verdict for the test case (line \ref{alg:57}-\ref{alg:571}). 
    
\item 
If the test verdict is not ``failed'', the possible modes 
of the CPS at the last moment of the simulation include ``final'' 
and the initial mode in the test case includes ``final'' or ``acceptable'', 
then the test oracle issues a ``passed'' verdict (lines \ref{alg:590}-\ref{alg:59}). 
    
\item 
If the test verdict is not ``failed'', and the possible modes of 
the CPS at the last moment of the simulation includes ``final'' 
and the initial mode in the test case include ``failing'', 
then the test oracle issues a ``failed'' verdict (line \ref{alg:60}).
    
\item 
If the test verdict is not ``failed'' and the possible modes 
of the CPS at the last moment of the simulation include ``acceptable'', 
then the test oracle issues a ``failed'' verdict (line \ref{alg:62}), 
because the CPS was not able to reach its target, i.e. the 
final mode or goal conditions, within the simulation time. 
Recall that the simulation time is the time within 
which the CPS must reach the goal condition. 
Otherwise, the CPS fails even if it can reach the goal in a longer time. 
In our inverted pendulum system, although the modes 
``max left'' and ``max right'' are both acceptable, 
the goal of designing such a CPS is to keep it in the 
``stabilized'' mode; therefore, if this CPS is not in 
the mode ``stabilized'' at the end of the simulation, 
the CPS failed to reach its goal. 
In this case, although the CPS is in an acceptable 
mode at the end of the simulation, {\sc HyTest} recognizes 
a failure since the goal was not reached. 

\item
If executing the simulation encounters any problem 
that leads to an exception, the test oracle 
issues a ``failed'' verdict, lines \ref{alg:63}-\ref{alg:631}.

\end{itemize}

\vspace*{-6pt}
\subsection{Implementation Details}

We implemented HyTest using Java 1.8.0 and Matlab R2022. 
Algorithms \ref{alg:conditionsGraphGen} and \ref{alg:testConditionGen} 
were implemented using Java and the rest of the implementation is in Matlab.
The Java code takes a hybrid model and 
other inputs, parses them, and generates a
condition graph and test conditions. 
The Matlab code obtains the test conditions and 
the rest of the inputs, generates and selects 
test cases, executes these test cases on the Simulink 
model of the CPS, and returns the test results.

\section{Empirical Study}
\label{sec:experiment}

\noindent To evaluate {\sc HyTest} we conducted an empirical 
study, asking the following research questions: \\

{\bf RQ1}: How effective is {\sc HyTest} at revealing faults in CPSs?

{\bf RQ2}: How efficient is {\sc HyTest} at testing CPSs?

\subsection{Objects of Study}

To conduct our study we require CPSs. 
{\sc HyTest} performs testing at the MiL level, 
and requires simulation models of our objects of study. 
To avoid possible threats to external validity that might
occur were we to design simulation models of our objects of 
study ourselves, we sought CPSs that came with simulation models. 
To find such CPSs we searched research monographs on related work.
Ultimately, this process led us to select five CPSs.\footnote{The
CPSs we selected are all available via links provided
in the relevant monographs, all of which are cited in this subsection.}
Table \ref{tab:objectsofstudy} provides data on the selected CPSs.

\begin{table}[t!]
\renewcommand\thetable{1}
\begin{center}    
\footnotesize
\caption{ Objects of Study} 
\label{tab:objectsofstudy}
\vspace*{-12pt}
\begin{tabular}{| c | c | c | c |}
\hline
& \# of Simulink  &  \# of Mutation & \# of Faulty  
\\
& Blocks  &  Operators  &  Models
\\
\hline
Cruise Control & 16 & 5 & 26
\\
\hline
Inverted Pendulum & 11 & 4 & 20
\\
\hline
Hexapod & 348 & 6 & 191
\\
\hline
Rooms and Heaters & 24 & 8 & 56 
\\
\hline
Automatic Transmission & 14 & 7 & 66
\\
\hline
\end{tabular}
\end{center}
\vspace*{-10pt}
\end{table}

``Cruise Control'' is a cruise control system \cite{CruiseControl}.
This system monitors the speed of a vehicle through 
sensors, increases and decreases the vehicle's 
speed to match a set speed, and maintains that speed. 
``Inverted Pendulum'' is the inverted pendulum system \cite{InvertedPendulum} 
introduced in Section \ref{subsec:CPS} and used 
to illustrate the operation of {\sc HyTest}. 
``Hexapod'' is a hexapod \cite{KhazaeeArticle} -- an 
autonomous legged robot that is able to move around its 
environment with high flexibility and stability.
This robot can reach a preset goal and 
avoid preset obstacles on its way to the goal. 
``Rooms and Heaters'' is a system \cite{FehnkerInPro} 
that controls transfer of moving heaters among 
adjacent rooms to maintain the temperature of 
the rooms within a desired range. 
Finally, ``Automatic Transmission'' is an automatic transmission 
system that controls an automobile's
speed and engine rpm \cite{ZhaoArticle}.
The simulation models for all five of these CPSs
are Simulink models, and they contain a wide range of 
block types including Integrator, Merge, Add, Transfer 
Function, If, State Chart, Relational and Logical Operators.
The Simulink models of these CPSs have each been 
designed using different types of controllers; this 
allows us to examine whether {\sc HyTest} can generate 
test cases for CPS's with different controllers.

For Inverted Pendulum we retrieved a hybrid 
model from \cite{INHybrid} and for Cruise Control 
we followed the description provided in 
\cite{CruiseControl} to recreate its model. 
The hybrid model for Hexapod is from \cite{KhazaeeArticle}. 
The hybrid models for Rooms and Heaters and Automatic 
Transmission are from \cite{AutomaticTransmissionHM1}
and \cite{AutomaticTransmissionHM2}.

To answer our research questions we require 
information on fault detection for our object CPSs. 
Unfortunately, no faults have been reported for these systems.
For this reason we turned to a mutation-based approach
for fault seeding, as has commonly been used in
prior empirical studies of testing techniques, including
techniques that operate on models
(see, e.g., \cite{ArrietaInPro2,ArrietaInPro3,MatinnejadInpro2}).

Matinnejad et al.~\cite{MatinnejadInPro} provide a comprehensive 
list of Simulink fault patterns based on experiences reported by
Delphi Engineers and from their review of the literature on CPSs. 
Based on this, they created a set of mutation operators for Simulink models.
To create a set of faulty models, we used these mutation
operators to inject mutations into the Simulink models 
for our objects of study, by applying each mutation operator 
to each of the blocks in the Simulink model to which it was applicable. 
Table~\ref{tab:objectsofstudy} provides statistics on the 
numbers of mutation operators we applied to the models 
(third column) and the number of faulty models obtained through 
this process (fourth column) for each of our objects of study. 

\vspace*{-6pt}
\subsection{Variables and Measures}
\label{subsec:vars}

\subsubsection{Independent Variable}

Our independent variable is the 
test case generation technique used.
We wished to compare {\sc HyTest} to state-of-the-art techniques.
Our survey \cite{SadriArticle} provides 
information on techniques that are available with 
implementations; we have also reviewed related 
techniques that are not covered in the survey. 
Among the techniques reviewed, only one is similar in aim 
and features (the basis of the technique, 
the type of testing it performs, and the simulation 
level it operates on) to {\sc HyTest}, 
and that is SimCoTest \cite{MatinnejadArticle1}.

To generate test cases, SimCoTest uses a meta-heuristic-based 
search algorithm that attempts to maximize diversity in 
test output signals that are generated by Simulink models. 
To address both continuous and discrete behaviors, SimCoTest 
generates test inputs as functions over time in Simulink 
models in an entirely black box manner. 
The technique provides manual test oracles that 
depend on on engineers' estimates of acceptable 
deviations from the expected results. 
To this end, SimCoTest needs a fault free 
version of a model and obtains the output 
signal of such a model (as ground truth).
SimCoTest validates/verifies CPSs at the MiL 
level while generating test cases for CPSs 
at the next levels (e.g., the SiL level), 
and it performs system testing. 
Although the authors of \cite{MatinnejadArticle1}
do not directly mention CPSs in their paper, 
that paper focuses on an extension of their previous work 
that targeted CPSs and used CPSs in its evaluation, 
so we have selected it as our primary baseline technique.

Several other techniques discussed in our survey 
\cite{SadriArticle} fail to be relevant as baselines because 
they are different in terms of simulation level, 
test level, and/or supported product. (Since these 
techniques are discussed in detail in the survey, 
we refer readers there for details.)
Several other techniques could potentially be relevant but 
were not provided with implementations.
In principle, we could attempt to implement such techniques, 
but we could not guarantee that our implementations 
would faithfully capture the proposed algorithms. 
Also, there are several techniques that generate 
test cases for testing CPSs that have been proposed  
recently and are not included in the survey, that focus on 
specific types of CPSs \cite{MoghadamInPro, KluckInPro, HumeniukInPro1,
FerdousInPro, StaraceInPro, CastellanoInPro, HumeniukInPro2, YanInPro}. 
For example, Moghadam et al. \cite{MoghadamInPro} 
propose a technique that generates critical test 
roads on which to test an autonomous-driving car.
Finally, as noted earlier, there are many test case
generation techniques that may be applicable to
CPSs that have not been presented, implemented,
or studied relative to CPSs; these are not
viable choices as baseline techniques in this case.

As a second baseline technique we utilized 
a random test case generation technique. 
While such a technique is not necessarily 
practical in practice, an algorithmic technique, 
if it is to be beneficial, should outperform 
a purely random technique.

\subsubsection{Dependent Variables}

To assess the effectiveness of techniques,
we calculate the percentage of faulty models in
which the techniques are able to detect faults.

To assess efficiency we calculate 
{\em total testing time}; this is 
the sum of the time required
to {\em generate and execute} the
test cases for a CPS.
In this study, we generate test cases only
on the hybrid model, so we report the time
required to do this as test case generation time.
We execute test cases, however, on each of
several faulty models, so we report
test execution time as the average of
the times required to execute test cases
over each of these models.

\vspace*{-6pt}
\subsection{Study Operation}
\label{sec:study}

For each of the five objects of study
HyTest first generated test cases, in the format discussed 
in Section \ref{sec:TCGen} and then extracted the 
test inputs and initial modes from the test cases. 
Next, HyTest put the CPS in different initial states 
by setting the model's initial values to the 
test inputs, provided the initial mode 
for the test oracle, and executed the faulty models. 
We recorded the numbers of generated test cases, 
the faulty models in which faults were detected,
the number of test executions, the test case 
generation time and the test execution time.

For SimCoTest, we determined input and output ranges. 
Also, the technique allows us to specify how much test case
generation time to allow, and a number of final test cases. 
We set the test case generation time to the time that 
{\sc HyTest} required to generate test cases, and 
the number of test cases to the number of test cases 
that {\sc HyTest} generated for each of our objects of study. 
Because SimCoTest generates test cases based on output 
diversity, where there is more that one output we 
needed to take that into account when specifying 
a number of test cases in order obtain the desired number. 
For example, the Rooms and Heaters CPS has three outputs. 
If {\sc HyTest} generates $n$ test cases for this CPS, 
then for SimCoTest we set the number of test cases to $ceil(n/3)$.

For the Random technique, we randomly generated exactly
as many random test cases as {\sc HyTest} generated 
to control for differences in numbers of test cases. 
To control for random elements of {\sc HyTest} and
the random test case generation technique, following
a recommendation in a similar study presented in \cite{HutchinsInPro}, 
we repeated the foregoing processes 30 times for each 
of the object systems; hence, the results reported in 
Tables \ref{tab:ResCC} to \ref{tab:ResAT} for 
{\sc HyTest} and Random are the averages of results over 30 runs.

\vspace*{-6pt}
\subsection{Threats to Validity}

{\em External validity} threats concern the generalization of our findings.
As objects of study we selected five CPSs, so our results pertain
to those CPSs and may not generalize beyond them.
As baseline techniques we chose SimCoTest and random test case generation.
Other techniques may compare differently.
To assess fault detection abilities we relied
on the insertion of mutations into Simulink models.
Our mutations are based on those defined in earlier
research in which they were derived based on experiences
from Delphi Engineers and other sources, so they
have some relevance to natural faults; nevertheless,
results obtained using these may not match results 
on natural faults occurring in practice.
The hybrid models obtained for our objects
represent only five specific models.
However, we did verify the simulation models against the 
hybrid models and specifications and made sure both models 
behaved in accordance with their corresponding specifications. 

{\em Internal validity} threats concern uncontrolled 
factors that may have affected our results.
Errors in our implementation of {\sc HyTest} could result 
in the collection of incorrect data; to reduce this 
threat we tested our implementation rigorously.
Randomness in the algorithms used by {\sc HyTest} and
random test case generation could also affect results.
To reduce this threat, we ran {\sc HyTest} and the
Random technique 30 times for each object. 

To allow {\sc HyTest} to operate on 
Simulink models, we modified them by 
adding input/output blocks to provide them
with test inputs (as initial values from the 
Matlab workspace) and to obtain the test 
outputs that the test oracle needs. 
Because such a modification might change the output 
of the models and result in incorrect results,
we reviewed the models' outputs to confirm that
the CPSs' behaviors were not affected.

{\em Construct validity} threats concern our metrics and measures. 
Our metric for efficiency does not account for other costs 
related to testing such as time spent debugging.

\vspace*{-6pt}
\subsection{Results}
\label{sec:results}

Tables \ref{tab:ResCC}, \ref{tab:ResIN}, \ref{tab:ResHX}, 
\ref{tab:ResRH}, and \ref{tab:ResAT} provide data for the 
techniques considered, for each of the 
five objects of study, respectively.
In each table, Column 1 lists the 
technique used to generate test cases,
Column 2 shows the number of test 
cases that were generated by the technique, 
and Column 3 shows the amount of time 
each technique required to generate those test cases. 
Column 4 shows the average time 
required to execute test cases on each faulty model 
using all generated test cases.
This column together with Column 3 present
the components of total testing time; these
are added to calculate total testing time, which
is presented in Column 5, and used to answer RQ2. 
Columns 6 and 7 show the number
and percentage of distinct faults that were revealed during 
the testing process; this data pertains to the 
effectiveness of techniquesm and used to answer RQ1.

Regarding RQ1, as Table \ref{tab:ResCC} shows, 
{\sc HyTest} was able to reveal all of the faults 
in Cruise Control while Random and SimCoTest revealed, 
respectively, 50\% and 65\% of the faults.
Tables \ref{tab:ResIN} and \ref{tab:ResHX} show
that on Inverted Pendulum and Hexapod, 
{\sc HyTest} was able to detect all of the faults, 
whereas Random and SimCoTest detected fewer than half. 
Table \ref{tab:ResRH} shows that on Rooms and Heaters,
{\sc HyTest} revealed all of the faults while Random 
and SimCoTest revealed just 54\% of the faults. 
As Table \ref{tab:ResAT} shows, {\sc HyTest} revealed 
all of the faults for Automatic Transmission, whereas
Random and SimCoTest revealed around one third of the faults. 
Hence, {\sc HyTest} was more effective at revealing 
faults than either of the baseline techniques on 
all of our objects of study

Regarding RQ2, data in Column 5 shows that 
{\sc HyTest} required less testing time than
the other two baseline techniques on all of 
our objects except Rooms and Heaters (see Table \ref{tab:ResRH}). 
On this CPS, Random required the least time to test the CPS 
and {\sc HyTest} required the second least time. 
It is clear from Column 3 that test case generation
time is the factor that causes this difference.
As noted in Section \ref{sec:TCGen}, {\sc HyTest} 
generates test cases in four major steps 
and in the third step it generates the input 
space using the Cartesian product operator,
Depending on how many variables the hybrid model has, 
how large the ranges of acceptable values for those 
variables are, and what precision they have, 
the time required to generate that Cartesian product 
varies and the input size changes, and this 
can directly affect test execution time.
Given the size of Hexapod in terms of variables,
this accounts for the differences in that time. 
As a partial answer to RQ2, based on the the 
results of our study, {\sc HyTest} was more efficient 
at revealing faults on four of our objects of study 
than either of the baseline techniques. 
We discuss this further in Section \ref{sec:discussion}.  

\begin{table}[t]
\begin{center}
\renewcommand\thetable{3.a}
\footnotesize
\caption{ Results for Cruise Control} 
\label{tab:ResCC}
\resizebox{\columnwidth}{!}{
\vspace*{-12pt}
\begin{tabular}{| c | c | c | c | c | c | c |}
\hline 
 & \# of &
  TC Gen &  Test Exe. &  Total Testing  &  \# of Faults & Pct. of Faults \\ 

   & TCs &
  Time  & Time&  Time  &  Revealed &  Revealed \\
  \hline  
 
HyTest & 11  & 0.053 m &  0.038 m  & 0.091 m  & 26 & 100\% \\ 
\hline 

Random  & 11 & 0.702 m & 0.043 m  & 0.745  & 13 & 50\% \\ 
\hline 
SimCoTest & 11 & 0.053 m &  0.040 m  & 0.093 m  & 17 & 65\% \\ \hline 
 \end{tabular}}
 \vspace*{-12pt}
 \end{center}
\end{table}

\begin{table}[t!]
\begin{center}
\renewcommand\thetable{3.b}
\footnotesize

\caption{ Results for Inverted Pendulum} 
\label{tab:ResIN}
\resizebox{\columnwidth}{!}{%
\vspace*{-12pt}
\begin{tabular}{| c | c | c | c | c | c | c |}
\hline 
 & \# of &
  TC Gen &  Test Exe. &  Total Testing  &  \# of Faults & Pct. of Faults \\ 

   & TCs &
  Time  & Time&  Time  &  Revealed &  Revealed \\
  \hline  

HyTest & 13 & 0.164 m & 0.048 m  & 0.212 m  & 20 & 100\% \\ 
\hline
Random  & 13 & 0.265 m & 0.051 m  & 0.316 m  & 9 & 45\% \\ 
\hline
SimCoTest & 14 & 0.164 m & 0.061 m  & 0.225 m  & 9 & 45\% \\ 
\hline
 \end{tabular}}
 \vspace*{-12pt}
 \end{center}
\end{table}

\begin{table}[t!]
\renewcommand\thetable{3.c}

\begin{center}   
\footnotesize

\caption{ Results for Hexapod} 
\label{tab:ResHX}
\resizebox{\columnwidth}{!}{%
\vspace*{-12pt}
\begin{tabular}{| c | c | c | c | c | c | c |}
\hline 
 & \# of &
  TC Gen &  Test Exe. &  Total Testing  &  \# of Faults & Pct. of Faults \\ 

   & TCs &
  Time  & Time&  Time  &  Revealed &  Revealed \\
  \hline

HyTest & 20  & 0.073 m & 0.601 m & 0.674 & 191 & 100\% \\ 
\hline  
Random  & 20 & 0.506 m & 0.659 m  & 1.165  & 80 & 42\% \\ 
\hline  
SimCoTest & 20 & 0.073 m & 1.086 m & 1.159  & 74 & 39\% \\ 
\hline  
 \end{tabular}}
 \vspace*{-12pt}
 \end{center}
\end{table}

\begin{table}[t!]
\renewcommand\thetable{3.d}

\begin{center}   
\footnotesize

\caption{ Results for Rooms and Heaters} 
\label{tab:ResRH}
\resizebox{\columnwidth}{!}{%
\vspace*{-12pt}
\begin{tabular}{| c | c | c | c | c | c | c |}
\hline 
 & \# of &
  TC Gen &  Test Exe. &  Total Testing  &  \# of Faults & Pct. of Faults \\ 

   & TCs &
  Time  & Time&  Time  &  Revealed &  Revealed \\
  \hline

HyTest & 43 & 11.462 m & 0.843 m  & 12.485 m  & 56 & 100\% \\ 
\hline
Random  & 43 & 0.332 m & 3.612 m & 3.944 m & 30 & 54\% \\ 
\hline
SimCoTest& 45 & 11.462 m & 2.609 m & 14.071 m  & 30 & 54\% \\ 
\hline
 \end{tabular}}
 \vspace*{-12pt}
 \end{center}
\end{table}

\begin{table}[t!]
\renewcommand\thetable{3.e}

\begin{center}
\scriptsize

\caption{ Results for Automatic Transmission} 

\label{tab:ResAT}
\resizebox{\columnwidth}{!}{%
\begin{tabular}{| c | c | c | c | c | c | c |}
\hline 
 & \# of &
  TC Gen &  Test Exe. &  Total Testing  &  \# of Faults & Pct. of Faults \\ 

   & TCs &
  Time  & Time&  Time  &  Revealed &  Revealed \\
  \hline  

HyTest & 15 & 2.068 m & 0.358 m & 2.426 m  & 66 & 100\% \\ 
\hline 
Random  & 15 & 0.365 m & 10.467 m & 10.832 m  & 22 & 33\% \\ 
\hline 
SimCoTest & 15 & 2.068 m & 7.512 m  & 9.580 m  & 21 & 32\% \\ 
\hline 
 \end{tabular}}
 \vspace*{-12pt}
 \end{center}
\end{table}

\section{Discussion}
\label{sec:discussion}

\noindent While we were applying {\sc HyTest} to the original, 
unmutated Simulink models for all systems, 
{\sc HyTest} revealed two additional faults 
that the baseline techniques did not reveal. 
One of these faults occurred because of a 
disallowed transition between modes of the 
CPSs, which may be the result of sudden changes in the values; 
the other was caused by changing from 
a failing mode to an acceptable mode, which 
means the CPSs continued to operate after they failed. 
We consider the latter a failure because the 
requirements of the object of study involved
included that it must halt when the CPS fails. 

As discussed in Section \ref{sec:results}, the test 
case generation time for Rooms and Heaters was greater than 
that for Random, primarily due to the time required
to generate the program's input space using a Cartesian product. 
This input space is required only if there are test conditions 
that are not covered by any test cases in the 
first two steps of test case generation. 
If all test conditions are 
covered in the first two steps, {\sc HyTest} 
will not proceed with the step that involves
the Cartesian product, and its test case
generation time will not include this expense.

In some cases it may be possible to set the initial 
values in a system's and CPS's dynamics to values that cover
all possible modes and transitions in the hybrid model. 
For example, in Hexapod, the test conditions that were 
not covered by the generated test cases at the first 
two levels were the ones that led the CPS to the failing mode. 
Hence, if we set the initial values on the system's dynamics 
to values that lead the system to a failure, then all the test conditions 
to the failing mode will be covered in the second step, 
which uses the system's dynamic to generate an input space. 
These values are easily determined by the test engineer or system designer.

The test case generation process using {\sc Hytest } 
is performed just once, immediately after the hybrid model of 
the CPS is ready and before any simulation model has been designed,
so the generation process can be run in parallel with
the work on the simulation model, and then may add no
real time to the overall development process.

Finally, since {\sc HyTest} was able to reveal all 
faults and they can be targeted and fixed at early 
stages of CPS development, i.e., at the MiL level, 
using {\sc HyTest} in these cases prevents the 
extra costs that would arise if those faults remained
hidden until later stages of development process.

\section{Related Work}
\label{sec:related}

\noindent Several papers (e.g., \cite{MatinnejadArticle1, MatinnejadInPro3, ArrietaInPro2,
MatinnejadInpro1, TurleaInPro, MatinnejadArticle, MatinnejadInpro2,
ZhaoArticle, LochauArticle, PohlheimArticle, ArrietaInPro3,
GadkariInPro, MenghiInPro, DeshmukhArticle}) present test case
generation techniques for CPS's using simulation models, 
assuming the models have been designed correctly, 
so they can be used as ground truth models for 
generating test cases that can be used at the SiL and HiL levels.
{\sc HyTest}, in contrast, uses hybrid models to generate test cases 
at the MiL level and to test simulation models of CPSs.

Several papers (e.g. \cite{OschInPRo,JuliusInPro,DangInPro}) 
present testing approaches that use models of 
hybrid systems without specifically considering CPSs. 
These approaches typically use hybrid models to 
generate test cases based on coverage to determine 
a systems' conformance with its specification.
{\sc HyTest}, in contrast, directly targets CPSs.
 
Several papers (e.g., \cite{BadbanInPro,BenderArticle}) ignore either 
the continuous or discrete dynamics of hybrid systems when generating test cases. 
Badban et al. \cite{BadbanInPro} 
present an approach that generates abstract test cases for hybrid 
systems as paths through an abstract representation model of the system. 
Discrete dynamics are not included in the 
definition of the system; this leads to testing just continuous dynamics of CPSs. 
Matinnejad et al.~\cite{MatinnejadInpro2} present an approach
that generates test cases for continuous controllers
but do not discuss controllers with mixed discrete-continuous behavior.
Bender et al. \cite{BenderArticle} propose an approach 
that generates test cases using the discrete behavior of a 
system without considering continuous behavior. 
{\sc HyTest}, in contrast, targets both continuous and 
discrete behaviors of CPSs by using their hybrid models 
to generate test cases and test those CPSs. 

Several approaches 
\cite{TanInPro, BrandlInPro, AichernigInPro, AICHERNIGArticle,LiArticle} 
use predefined sets of behaviors 
(either correct or faulty) to generate test cases. 
These approaches restrict the test case generation 
process to behaviors that are known beforehand and ignore 
behaviors that the target system may exhibit under 
conditions not considered.
Tan et al. \cite{TanInPro} provide a test case
generation approach that simulates the behavior 
of a system in its environment and generates 
test cases from the simulation trace. 
They then create a testing automaton for each test case, 
which supplies the test case during the execution of 
the system model and later on its implementation. 
Approaches proposed in \cite{BrandlInPro, 
AichernigInPro, AICHERNIGArticle} to test the 
conformance of hybrid systems' implementations with 
specifications use qualitative reasoning 
and model checking to generate test cases. 
They mutate the model of the system's specification 
to obtain faulty system behaviors, and 
generate traces that are used to create test cases. 
This limits testing to the set of mutation operators. 
{\sc HyTest}, in contrast, 
is not restricted to known behaviors and applies to
all predefined and non-determined behaviors of CPSs.

Several approaches \cite{KalajiInPro, ChengInPro,
LiInPro, GorenArticle, SantiagoInPro,WangInPro} 
generate test cases using extended finite state 
machines (EFSM) or their variations, but contain
no discussion about applications to CPSs. 
The primary difference between hybrid models and extended 
finite state machines is that in a hybrid model there
is no start state, while in EFSMs there are start states.
Also, EFSMs use a set of input symbols 
\cite{ChengInPro1} whereas hybrid models do not.
Therefore, we believe that additional work is required 
to determine whether these approaches are applicable to,
and are effective and efficient, in testing CPSs. 
{\sc HyTest}, in contrast, directly targets CPSs.

There are several approaches that generate test cases for specific 
types of CPSs \cite{MoghadamInPro, KluckInPro, HumeniukInPro1, 
FerdousInPro, StaraceInPro, CastellanoInPro, HumeniukInPro2, YanInPro}. 
For example, Moghadam et al. \cite{MoghadamInPro} propose an approach 
that generates critical test roads on which to test an autonomous-driving car. 
{\sc HyTest}, in contrast is not limited to specific types of CPSs.

Several approaches, such as those presented in 
\cite{MenghiInPro} and \cite{AnnpureddyInPro}, 
focus on falsifying test cases and ignore passing ones. 
These approaches do not check whether 
the CPS functions correctly as specified.
For example, Menghi et al. \cite{MenghiInPro} 
propose an approach that generates test 
cases using falsification techniques, 
that generates an approximation of the 
system model and tries to find a falsifying input, 
i.e., an input that violates the requirement(s). 
{\sc HyTest}, in contrast, generates test cases that lead 
the CPS to failures and test cases that pass 
in order to examine both passing and failing 
behaviors.

Aerts et al. \cite{ReniersInPro} propose an approach 
for generating conformance test cases using hybrid models of CPSs. 
Their approach assumes that guards are not 
time-dependent, which renders
the approach non-applicable to CPSs modeled using timed automata. 
{\sc HyTest} does not have such restrictions. 

Several test case generation approaches for CPSs (e.g., \cite{MatinnejadArticle1,
ArrietaInPro2, MatinnejadArticle, MatinnejadInpro2, LochauArticle,
PohlheimArticle, ArrietaInPro3, GadkariInPro, MenghiInPro,
DeshmukhArticle, AlenaziInPro, GorenArticle, AraujoInPro,
CukicArticle, CarterInPro, MaArticle, KuroiwaArticle, SchneiderInPro, 
ZhangArticle, SinhaInPRo, ReniersInPro}  
generate redundant test cases; these lead to unnecessary
test executions and unneeded testing costs. 
In contrast, {\sc HyTest} assumes that test cases from 
the same partitions in input spaces are redundant and 
does not execute all of them.

Several approaches for testing CPSs 
(e.g., \cite{TurleaInPro, ZhaoArticle, PohlheimArticle, GadkariInPro,
AlenaziInPro, GorenArticle, CukicArticle, SchneiderInPro, ZhangArticle,LiArticle}) rely 
on manual test oracles or provide a test oracle that compares the 
test outputs against a ground truth. 
These test oracles are similar to traditional test oracles used to test non-reactive systems, 
whereas CPSs are reactive systems and need oracles that 
check the correctness of sequences of interactions with the environment. 
In CPSs, a standard correctness requirement 
is that all executions be allowed based on the system 
requirements and specifications. Checking the correctness 
of the final results  against a ground truth may indicate 
a ``failure'' when the test results are not equal to 
the ground truth, whereas the CPS may have interacted 
with its environment acceptably and reached its goal. 
{\sc HyTest} supports the testing of reactive systems.

\vspace*{-6pt}
\section{Conclusion}
\label{sec:conclusion}

\noindent We have presented {\sc HyTest}, an approach for 
generating test cases for CPSs, that is accompanied by a test oracle. 
To evaluate the effectiveness and efficiency of {\sc HyTest} we 
studied its application to a set of CPSs, and compared its results 
to those of two baseline techniques: Random test case generation and SimCoTest. 
Our results show that {\sc HyTest}, with its test oracle, 
was able to expose more faults than the baseline 
techniques either in less time, or in a practically
insignificant greater amount of time.

Flaky test cases are test cases that are non-determistic
in terms of their results, possibly due to uncertainties 
and changing conditions in a system. As mentioned earlier, 
{\sc HyTest} captures all possible conditions that the 
values of the variables of a CPS may encounter during 
its execution. Therefore, as future work, we intend to
extend {\sc HyTest} to identify potential flaky tests.

We have studied {\sc HyTest} in relation to the testing of
CPSs at the MiL level; however, we believe that the test cases 
thus generated could be effective at the SiL and HiL levels. 
Additional studies could be performed to assess this.

In our empirical study, we evaluated {\sc HyTest} on five 
CPSs of various (low, medium, high) complexity and our results showed 
that {\sc HyTest} was effective and efficient when testing them. 
Although we do not have any theoretical or practical 
evidence that {\sc HyTest}'s applicability and 
scalability might be limited on larger and more complex
industrial CPSs, additional empirical work is needed to 
determine this.

\begingroup
\raggedright
\bibliographystyle{IEEEtran}
\bibliography{main}{}
\endgroup

\end{document}